\documentclass[12pt,a4paper,twoside]{article}
\usepackage{epsfig}
\begin{document}
\newcommand{\ttbar}     {$t\overline{t}$~}
\newcommand{\ppbar}     {$p\overline{p}$~}
\newcommand{\qqbar}     {$q\overline{q}$~}
\newcommand{\MET}       {\mbox{$\not\!\!E_T$}~}
\newcommand{\METns}     {\mbox{$\not\!\!E_T$}}
\newcommand{\pt}        {$p_T$~}

\title{Top Quark Physics at the Tevatron
\thanks{Presented at the ``XXVII Physics in Collision'' Symposium,
  Annecy, France, \mbox{26 - 29 June 2007.}}%
}
\author{Marc-Andr\'e Pleier\\
    \small Physikalisches Institut, Universit{\"a}t Bonn\\
    \small Nussallee 12, 53115 Bonn, Germany\\
    \normalsize On behalf of the CDF and D0 Collaborations}
\date{} %no date!
\maketitle
\begin{abstract}
  The Tevatron proton-antiproton collider at Fermilab with its centre
  of mass energy of 1.96 TeV is currently the only source for the
  production of top quarks. Its increased luminosity and centre of
  mass energy in Run~II allow both collider detectors CDF and D0 to
  study top quarks with unprecedented scrutiny.\\
  Recent results on the top quark's pair production cross section and
  its properties such as mass, electric charge, helicity of the $W$
  boson in its decay and branching fraction B($t \rightarrow Wb$) are
  presented and probe the validity of the Standard Model.
 \end{abstract}
%14.65.Ha 	Top quarks
%12.38.Qk 	Experimental tests
\section{Introduction}
The existence of the top quark as the weak isospin partner of the
bottom quark was already discussed in 1977 with the discovery of the
bottom quark and hence a third quark generation. The self consistency
of the Standard Model (SM) both required the existence of the top quark and allowed for
increasingly precise predictions of properties like its mass from
electroweak precision measurements. The top quark finally was
discovered in 1995 by the CDF and D0
collaborations~\cite{top_discovery} in the mass range predicted by
the SM, completing the SM quark sector. \\
With a mass of 170.9 $\pm$ 1.8 GeV~\cite{topmass}, the top quark
exhibits both the largest and the most precisely measured quark mass.
It is the fermion coupling most strongly to the Higgs boson, and due to its extremely
short lifetime of $\approx 4\cdot 10^{-25}$~s, it is the only quark that
decays before it can hadronise, allowing the study of the decay of an
essentially free quark.\\
Measuring the production cross section of the top quark and its
different properties such as mass, electric charge, $W$ boson helicity
in its decay, branching fraction B($t \rightarrow Wb$), {\it etc.},
and comparing with predictions of the SM is a very powerful tool in
searching for physics beyond the SM. The recently recorded large
datasets at the Run~II Tevatron allow for never-before-performed measurements of top quark
properties like the electric charge,
while other measurements like the top mass reach such a precision
that they start to become limited by systematic uncertainties.\\
This article focusses on the production of top quarks in pairs via the
strong interaction using datasets with an integrated luminosity of up
to 1~fb$^{-1}$. It also includes some measurements that became available
shortly after the conference date. The production of single top quarks
via the electroweak interaction with a first evidence observed by D0
is discussed in a different article in these proceedings
\cite{STtalk}. A detailed description of the CDF and D0 detectors can
be found in \cite{detectorCDF, detectorD0}.
%Some of the measurements presented in these proceedings became
%available shortly after the conference date.
%1~fb$^{-1}$ 
\section{Top Quark Pair Production and Decay}
In \ppbar collisions at a centre of mass energy $\sqrt{s}=1.96$ TeV,
top quarks are produced predominantly in pairs: \ppbar $\rightarrow$
\ttbar + X via the strong interaction (85\% \qqbar annihilation and
15\% gluon-gluon fusion). At next-to-next-to-leading order, the
corresponding SM cross section is 6.77 $\pm$ 0.42 pb for a top quark
mass of 175 GeV~\cite{ttbar_xsec}. According to the SM, the top quark
decays predominantly into $W$ bosons and $b$-quarks. Hence, there are
three event classes to be observed resulting from \ttbar decay, which
depend on the decay mode of the $W$ bosons: \\
{\bf (i)} a so-called {\it dilepton} final state where both $W$ bosons decay
leptonically, resulting in two isolated high-\pt leptons, missing
transverse energy \MET corresponding to the two neutrinos and two
jets, {\bf (ii)} a {\it lepton+jets} final state where one $W$ boson
decays leptonically, the other one hadronically, resulting in one
isolated high-\pt lepton, \MET and four jets, and {\bf (iii)} an {\it
  all-hadronic} final state where both $W$ bosons decay to
$\overline{q}q'$ pairs, producing six jets. In all final states, two of
the jets are $b$-jets (originating from the hadronisation of a 
$b$-quark). Additional jets can arise from initial and final state
radiation.\\
The all-hadronic final state represents the biggest branching fraction
of \ttbar events ($\approx$46\%), but it is also difficult to separate
from the large multijet background. The dilepton final state not counting
$\tau$ leptons constitutes $\approx$5\% of the \ttbar events and gives
the cleanest signal but suffers from low statistics. The
lepton+jets events in the $e$+jets or $\mu$+jets channels yield
$\approx$29\% of the branching fraction and provide the best
compromise between sample purity and statistics.
\section{Top Quark Pair Production Cross Section Measurements}
Measurements of the top quark pair production cross section provide an
important test of the predictions from perturbative QCD calculations
at high transverse momenta. Analysing different decay channels helps
to improve statistics of top events and studies of properties, as
well as the probing of physics beyond the SM that might result in
enhancement/depletion in some particular channel via novel production
mechanisms or decay modes. Instead of quoting single cross section 
results in the following subsections, all current measurements are
summarised in Figure~\ref{fig:xsec}.\\
Due to the selection of datasets enriched
in top quark pairs and the necessity of understanding object
identification, background modelling and sample composition, cross
section measurements form the foundation for all further property
analyses like the ones
% The
% datasets characterised by the cross section measurements serve as a
% starting point for the measurements of top quark properties
described in the subsequent sections of this article.
\subsection{Dilepton Final State}
Dilepton events are usually selected by requiring two isolated high
\pt leptons, \MET and at least two central energetic jets in an event.
The main physics backgrounds exhibiting both real leptons and \MET
arise from Z/$\gamma^{*}$+jets production with $Z/\gamma^{*}\to\tau^{+}\tau^{-},
\tau\to e,\mu$ and the production of dibosons ($WW,ZZ,WZ$).
Instrumental backgrounds mainly arise from misreconstructed \MET due to
resolution effects in Z/$\gamma^{*}$+jets production with $Z/\gamma^{*}\to
e^{+}e^{-}/\mu^{+}\mu^{-}$, but also from $W$+jets and QCD multijet
production where one or more jets fake the isolated lepton signature.
While the physics backgrounds are estimated from Monte Carlo, the
instrumental backgrounds are usually \mbox{modelled} using data.\\
The selected data samples can be further enriched in signal by
requiring additional kinematical event properties like the scalar sum
of the jet $p_T$s $H_{T}$ to be above a certain threshold or
discarding events where both leptons have the same electric charge.
The purities in these samples are usually quite good with a signal to
background ratio (S/B) better
than 2 at least, but the statistics are low. To enhance the acceptance
for dilepton final states and especially include ``1~prong'' hadronic
$\tau$ decays, the selection can be loosened to require only one fully
reconstructed isolated lepton ($e,\mu$) in addition to an isolated track
(``$\ell$+track analysis'').\\
A first measurement of the lepton+tau final state was recently
performed by D0, using events with hadronically decaying isolated taus
and one isolated high \pt electron or muon. In this case, the sample
purity was enhanced by requiring $b$-jet identification in the event.
The result is shown together with other measurements in
Figure~\ref{fig:xsec}.
\subsection{Lepton+Jets Final State}
Lepton+jets events typically are selected by requiring one isolated
high \pt lepton ($e,\mu$), \MET and at least 4 jets. The main physics
background here comes from $W$+jets production while the main
instrumental background arises from QCD multijet production where a
jet fakes the isolated lepton signature. Accordingly selected samples
exhibit a S/B around 1/2.\\
It is possible to extract the cross section using either purely
topological and kinematic event properties combined in a multivariate
discriminant to separate the \ttbar signal from background, or by
adding identification of $b$-jets. An advantage of topological
analyses is that they do not depend on the assumption of 100\%
branching of $t \rightarrow Wb$ and are therefore less
model-dependent than tagging analyses. Requiring $b$-jet
identification, however, is a very powerful tool to further suppress the
backgrounds which typically exhibit little heavy flavour content,
allowing for signal extraction in lower jet multiplicities and
providing very pure signal samples: a S/B $>$ 10 can easily be
achieved in selections requiring at least four jets if two
identified b-jets are required.\\
$b$-jets can be identified based on the long {\it lifetime} of B
hadrons resulting in significantly displaced secondary vertices with
respect to the primary event vertex or large significant impact
parameters of the corresponding tracks. Combining this type of information
into a neural network tagging algorithm, $b$-tagging efficiencies of
about 54\% are achieved while only about 1\% of light quark jets are
misidentified as $b$-jets, resulting in an improved S/B in tagged analyses. A second
way to identify $b$-jets is to reconstruct {\it soft leptons} inside a
jet that come from semileptonic B decays -- so far only
soft-$\mu$ tagging has been deployed in \ttbar analyses.\\
Using lifetime $b$-tagging, D0 was also able to perform a first
$\tau$+jets cross section analysis, using events with hadronically
decaying isolated taus -- the result is shown together with other
measurements in Figure~\ref{fig:xsec}.
\subsection{All-Hadronic Final State}
To study the all-hadronic final state, events with at least six
central energetic jets and no isolated high \pt leptons are selected. 
Since QCD multijet production represents a background which is orders of
magnitude larger than the signal process, $b$-jet identification is
mandatory for this final state. To allow further
separation of signal and background, multivariate discriminants based
on topological and kinematic event properties are deployed.
\subsection{Summary}
\begin{figure}[ht]
  \centering
  \includegraphics[width=0.42\textwidth]{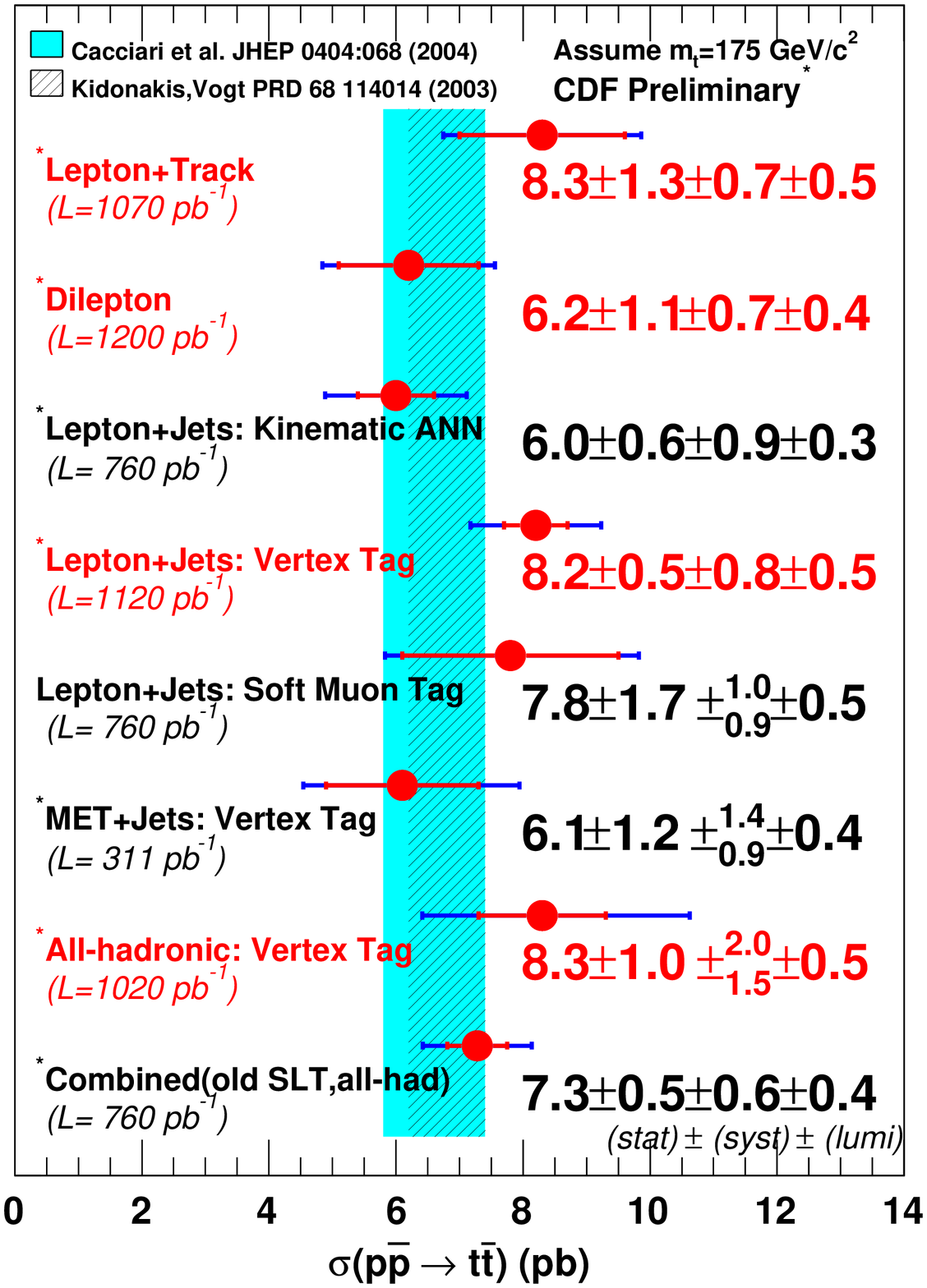}
  \includegraphics[width=0.5\textwidth]{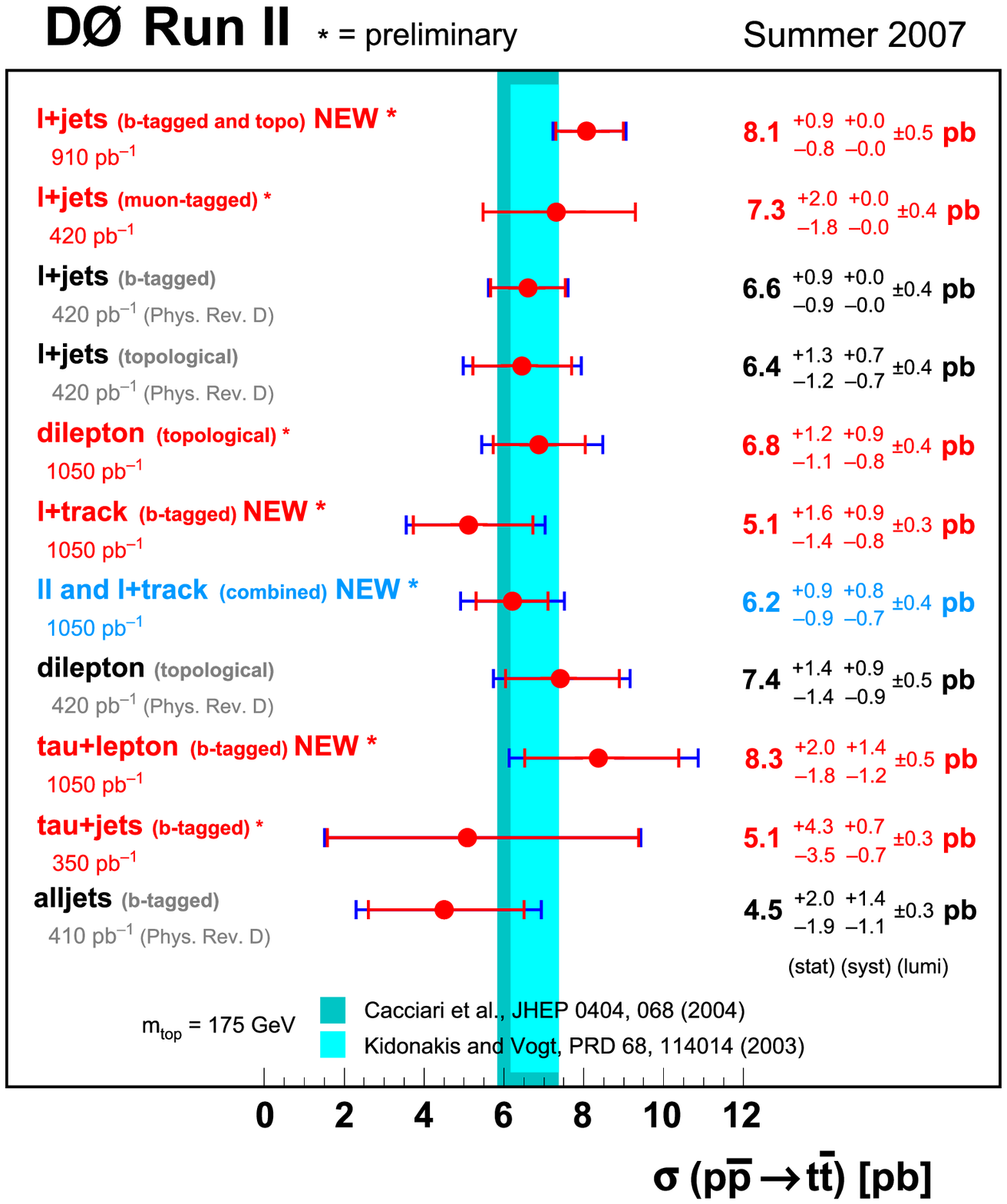}
  \caption{Top quark pair production cross-section measurements by CDF
    (left) and D0 (right) in comparison with theory predictions shown
    as coloured bands. If a systematic uncertainty is shown as 0,
    it is included in the statistical uncertainty.}
  \label{fig:xsec}
\end{figure}
Figure~\ref{fig:xsec} shows an overview of recent cross section
measurements performed by CDF and D0. All measurements
are in good agreement with the SM prediction and with each other -- the
single best measurements reaching a relative precision of
$\Delta\sigma/\sigma$=12\%. 
With increasing datasets, these measurements naturally start
to become limited by systematic uncertainties which in return can
be further constrained using additional data. For an integrated
luminosity of 2~$fb^{-1}$, a relative precision of
$\Delta\sigma/\sigma$=10\% should be achievable, providing stringent
tests to theory predictions.
\section{Measurement of B($t \rightarrow Wb$) / B($t \rightarrow Wq$)}
The ratio of branching fractions R = B($t \rightarrow Wb$) /
$\Sigma_{q=d,s,b}$ B($t \rightarrow Wq$) is constrained within the SM
to $0.9980 < R < 0.9984$ at 90\% CL~\cite{PDG}, assuming three fermion
generations, unitarity of the CKM matrix and neglect of non-$W$ boson
decays of the top quark. The most precise measurement to date has been
performed by D0 in the lepton+jets channel using data corresponding to
an integrated luminosity of 900~pb$^{-1}$ by comparing the event
yields with 0, 1 and 2 or more $b$-tagged jets. This measurement
obtains the following result for R from a simultaneous fit of R and
the \ttbar cross section (which is also shown in
Figure~\ref{fig:xsec}):
R = 0.991$^{+0.094}_{-0.085}\:{\rm(stat+syst).}$
% \begin{eqnarray}
%   R = 0.991^{+0.094}_{-0.085}\:{\rm(stat+syst).}\nonumber
% \end{eqnarray}
This result is in agreement with the SM expectation.
\section{Measurement of the Top Quark Electric Charge}
The electric charge of the top quark can be inferred from the electric
charges of its decay products. However, there is an inherent ambiguity
when pairing $W$ bosons and $b$-jets in a top quark pair event
resulting in possible charges of $|Q|$= 2e/3 or 4e/3, the latter being
predicted in exotic models~\cite{exotictopq}. D0 has published a first
measurement of the electric charge of the top quark excluding the
hypothesis of only exotic quarks of charge $|Q|$= 4e/3 being produced
at 92\% CL by analysing lepton+jets events with two identified
$b$-jets, obtaining the $W$ boson charge from the lepton charge, the
$b$-jet charge from a track-based jet charge algorithm and assigning the
$W$ boson -- $b$-jet pairing based on a kinematic event
fit~\cite{d0topq}. Similarly, CDF obtained a preliminary result on the
top quark charge using lepton+jets and dilepton events with an
observed $2\ln$(Bayes Factor) of 12.01, meaning that the data favour
very strongly the SM top quark hypothesis over the exotic model.
\section{Measurement of the $W$ Boson Helicity in $\rm t\bar{t}$ Decays}
Top quark decay in the V$-$A charged current weak interaction proceeds
only via a left-handed ($f^{-}$= 30\%) and a longitudinal
($f^{0}$=70\%) fraction of $W$ boson helicities, which is reflected in
the angular distribution of the charged lepton relative to the line of
flight of the top quark in the $W$ boson rest frame in lepton+jets
final states. Any observed right-handed fraction $f^{+} >
\mathcal{O}(10^{-4})$ would indicate physics beyond the SM.\\ The most
precise measurement to date has been performed by CDF using a dataset
corresponding to an integrated luminosity of 1.7 fb$^{-1}$ and
comparing the above mentioned angular distribution in data to
templates for longitudinal, right- and left-handed signal plus a
background template. When fitting both $f^{0}$ and $f^{+}$
simultaneously, the result is $f^{0} = 0.61 \pm 0.20\:{\rm(stat)} \pm
0.03\:{\rm(syst)}$, $f^{+} = -0.02 \pm 0.08\:{\rm(stat)} \pm 0.03
\:{\rm(syst)}$; constraining $f^{0}$ respectively $f^{+}$ to their SM
values when fitting $f^{+}$ respectively $f^{0}$, the result is $f^{0}
= 0.57 \pm 0.11\:{\rm(stat)} \pm 0.04\:{\rm(syst)}$, $f^{+} = -0.04
\pm 0.04\:{\rm(stat)} \pm 0.03 \:{\rm(syst)}$, $f^+<0.07~{\rm
(95\%~CL)}$, in agreement with expectations from the SM.
\section{Measurement of the Top Quark Mass}
\begin{figure}[h]
  \centering
  \includegraphics[width=0.5\textwidth,clip=]{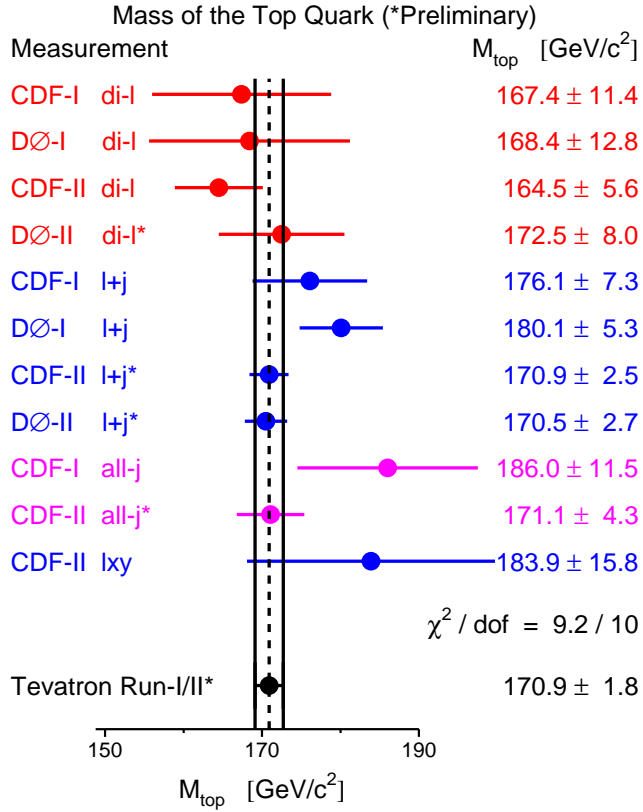}
  \caption{$m_{t}$ measurements used as an input for the
    current preliminary world average.}
  \label{fig:topmass}
\end{figure}

The top quark mass is a fundamental SM parameter that is not predicted
by the SM theory itself. It can be used in conjunction with the $W$ boson
mass to constrain the mass of the still undiscovered Higgs boson via
radiative corrections.\\
One of the most crucial ingredients for top quark mass measurements
%resulting in the dominant systematic uncertainty 
is the jet energy scale (JES), relating the measured jet energy to the
parton energy. Top events can provide an additional in situ calibration source
via hadronic $W$ boson decays by using the well-known mass of the $W$
boson as a constraint. Additional future constraints on the $b$-JES
could be derived from the study of $Z\to b\bar{b}$ decays.\\ The most
precise measurements of the top quark mass are achieved in the
lepton+jets final state due to the high branching fraction and yet
good S/B, the hadronic $W$ boson decay allowing for additional JES
calibration and the overconstrained event
kinematics as only one neutrino is present in the final state.\\
The top quark mass has been measured in all decay modes using
different analysis methods that can be roughly separated in two
categories: {\it Template methods} compare distributions of
observables sensitive to the top quark mass in data with template
distributions for varying top quark masses. {\it Dynamical methods}
try to maximise the use of information in each candidate event by
calculating a per-event probability density
% for being signal respectively background
as a function of the top quark mass, usually based on leading order
matrix elements.\\ An example for a dynamical method is the ``Matrix
Element Method'' pioneered by D0 in the lepton+jets channel during
Run~I which yielded the most precise single measurement of that
time~\cite{d0memassrunI}. It also gives the most precise
measurements obtained thus far in Run~II for both CDF and D0 in the lepton+jets
channel. Using the measured fraction of signal events, in these
measurements the per-event probability is obtained as a linear
combination of the signal and background probabilities. These are
evaluated based on leading order matrix elements for the $t\bar{t}$
signal and $W$+jets main background, also folding in the detector
resolution and summing over the different possible jet-parton
assignments, solutions for the longitudinal neutrino momentum and
possibly $b$-tagging event probabilities.\\
Fig. \ref{fig:topmass} shows the results of these measurements
together with all other measurements considered for the current world
average mass of 170.9~$\pm$~1.8~GeV~\cite{topmass}. By the end of
Run~II, a final top quark mass uncertainty of $\Delta m_{t}= 1$~GeV
should be achievable, which will -- together with an improved
measurement of the $W$ boson mass -- provide stringent constraints on
the mass of the Higgs boson. Following from the current world average
top quark mass, the most likely Higgs boson mass is $m_{H}=76^{+33}_{-24}$
GeV respectively $m_{H}< 144$~GeV at 95\% CL, promising an interesting
competition between Tevatron and LHC for the first evidence of the
Higgs boson.
\section{Summary}
A wealth of top analyses is being pursued at the Tevatron, which continue
to probe the validity of the SM. While some measurements are reaching
the precision of the theory predictions and thus provide stringent
tests thereof, others are still statistically limited, leaving room
for physics beyond the SM. So far, all measurements are in agreement
with the SM. More detailed descriptions of the analyses presented here
can be found online~\cite{top_group_web}.\\ Continuously improving the
analysis methods, and using the increasing integrated luminosity from
a smoothly running Tevatron, expected to deliver more than 6~fb$^{-1}$
by the end of Run~II, we are moving towards precision measurements and
hopefully discoveries within and outside the SM.
\section*{Acknowledgements}
The author would like to thank the organisers for a rich and rewarding
conference and to acknowledge the generous support of the Alexander
von Humboldt Foundation.


\begin{thebibliography}{99}
\bibitem{top_discovery}
  CDF Collaboration, F. Abe {\it et al.}, Phys. Rev. Lett. {\bf 74}, 2626 (1995);\\
  D0 Collaboration, S. Abachi {\it et al.}, Phys. Rev. Lett. {\bf 74}, 2632 (1995).
\bibitem{topmass} Tevatron Electroweak Working Group,
  \verb|hep-ex/0703034| (2007).
\bibitem {STtalk} C.~Ciobanu, for the CDF and D0 collaborations, ``Single-Top-Quark Physics
   at Hadron Colliders'', these proceedings.
%\bibitem {BSMtalk} J.~Nachtman , for the CDF and D0 collaborations, ``Collider Searches for Physics Beyond the Standard Model'', these proceedings.
\bibitem {detectorCDF} CDF Collaboration, R. Blair {\it et al.},
  % The CDF-II Detector: Technical Design Report, 
  Fermilab-Pub-96-390-E (1996).
\bibitem {detectorD0}  D0 Collaboration, V.~M.~Abazov {\it et al.},
  % ``The upgraded D0 detector,''
  Nucl.\ Instr.\ Meth.\  A {\bf 565}, 463 (2006).
  % [arXiv:physics/0507191].
\bibitem{ttbar_xsec}
  N.~Kidonakis and R.~Vogt,
  %``Next-to-next-to-leading order soft-gluon corrections in top quark
  %hadroproduction,''
  Phys.\ Rev.\ D {\bf 68}, 114014 (2003).
\bibitem{PDG}  S. Eidelman {\it et al}, Phys. Lett. B {\bf 592}, 1 (2004).
%the Vtb limits are actually only in the older version...
%W. M. Yao {\it et al.}, Journal of Physics G {\bf 33}, 1 (2006).
\bibitem{exotictopq} D. Chang {\it et al.}, Phys. Rev. D {\bf 59},
091503 (1999); {\bf 61}, 037301 (2000);\\ D. Choudhury {\it et al.},
Phys. Rev. D {\bf 65}, 053002 (2002).
\bibitem{d0topq} D0 Collaboration, V.~M.~Abazov {\it et al.},
Phys. Rev. Lett. {\bf 98}, 041801 (2007). %PRL
\bibitem{d0memassrunI} D0 Collaboration, V.~M.~Abazov {\it et al.},
  Nature {\bf 429}, 638 (2004).
\bibitem{top_group_web}
  \verb|http://www-d0.fnal.gov/Run2Physics/top/index.html|;
  \verb|http://www-cdf.fnal.gov/physics/new/top/top.html|
\end{thebibliography}
\end{document}